\documentclass[12pt]{iopart}

\usepackage[dvips]{graphicx}

\newcommand{\cm}{cm$^{-1}$}

\begin{document}

\title[Calculations of Yb$_2$]{Accurate calculations of 
the dissociation energy, equilibrium distance and
spectroscopic constants for the Yb dimer}

\author{N S Mosyagin$^1$\footnote{http://www.qchem.pnpi.spb.ru}, 
A N Petrov$^{1,2}$, A V Titov$^{1,2}$} 

\address{$^1$Petersburg Nuclear Physics Institute, 
             Gatchina, St.-Petersburg district 188300, Russia}
\address{$^2$St.-Petersburg State University, St.-Petersburg, Russia}

\ead{mosyagin@pnpi.spb.ru}

\begin{abstract}
 The dissociation energy, equilibrium distance, and spectroscopic constants 
 for the $^1\Sigma_g^+$
 ground state of the Yb$_2$ molecule are calculated.  The relativistic effects
 are introduced through generalized relativistic effective core potentials 
 with very high precision. The scalar relativistic coupled cluster method
 particularly well suited for closed-shell van-der-Waals systems is used for
 the correlation treatment. Extensive generalized correlation basis sets were
 constructed and employed. The relatively small
 corrections for high-order cluster amplitudes
 and spin-orbit interactions are taken into account
 using smaller basis sets and the spin-orbit density functional theory.
\end{abstract}


\submitto{\jpb}

\maketitle

\section{Introduction.}

 Several groups are working on the development of a new generation of
 frequency and time standards based on atomic optical
 transitions\cite{Bergquist:01, Diddams:04, Takamoto:05}.  Neutral ytterbium
 atoms loaded into a laser lattice are very promising candidates for 
 constructing such high-performance atomic clock\cite{Hoyt:05,Barber:08}.  
 The properties
 of Yb$_2$ are necessary to assess the feasibility of using laser cooled 
 and trapped Yb atomic species for ultraprecise optical clocks or quantum
 information-processing devices\cite{Kotochigova:03}.

 Unfortunately, reliable experimental data on the 
 dissociation energy, equilibrium distance, and 
 spectroscopic constants of the Yb$_2$ molecule are unknown; e.g., 
 the uncertainty of the experimental dissociation energy 
 estimate from \cite{Guido:72}, 0.17~eV, is comparable 
 to the value itself. A series of papers was devoted to their calculation. 
 Dolg and co-workers reported the values 400 \cite{Dolg:92a},
 470 \cite{Wang:98b}, and 740~\cm \cite{Cao:02} for the dissociation energy. 
 The coupled electron pair approximation, density
 functional theory (DFT), configuration interaction method with single and
 double excitations (CISD), and coupled cluster method with single, double
 (and non-iterative triple) cluster amplitudes, CCSD(T), were used to account
 for the correlation effects. Ytterbium core shells were replaced by scalar
 (spin-averaged) energy-consistent pseudopotentials (PPs) generated for 2, 10,
 and 42 explicitly treated electrons. Their latest result\cite{Cao:02} of
 740~\cm~should be considered the most reliable, because the basis set used was
 larger than that in the previous calculations and 42 electrons were 
 treated explicitly for each Yb atom (42e-PP).
 
 In \cite{Wu:04}, the scalar DFT approach and a 24-electron relativistic
 effective core potential (RECP) model for ytterbium were used. The 
 dissociation
 energy estimates ranging from 500 to 1400~\cm~were obtained with different
 exchange-correlation functionals. In \cite{Buchachenko:07}, the Yb
 dimer was studied within the averaged quadratic coupled cluster,
 CCSD(T), and DFT
 approximations. The Yb atom was described by a 42-electron energy-consistent 
 PP. It was emphasized that the ``incomplete convergency, most clearly seen 
 for Yb$_2$ results, indicated the need for more advanced ab initio schemes''. 
 In addition to the evident problem of the incompleteness of the one-electron 
 basis set, it is not clear
 whether the truncation of the cluster expansion after the three-body terms
 provides a good approximation for the Yb$_2$ ground state, which could be 
 considered as a perturbed four-electron system. 
 Furthermore, almost all calculations mentioned above were done within
 the scalar relativistic approximation. In spite of the closed-shell-like
 nature of the system under study, the contribution of spin-dependent
 interactions to the bond energy can still be significant (cf.\
 \cite{Petrov:09a}). The only attempt\cite{Dolg:92a} to estimate the role of 
 spin-orbit interactions
 in Yb$_2$ was made within a somewhat simplistic four-electron CI scheme using
 a very restricted basis set
 that can hardly be used to reliably reproduce the van-der-Waals behaviour of 
 the potential curve.

 In the present paper, we report our results of
 improving the accuracy of the calculated
 dissociation energy, equilibrium distance and spectroscopic
 constants for Yb$_2$ using extremely flexible generalized
 correlation basis sets, contributions from high-order cluster amplitudes and
 spin-dependent relativistic effects.
 Such improvements were successfully applied to accurate calculation of
 Hg$_2$ when giving a few times better agreement with the experimental data
 than the other earlier performed studies \cite{Petrov:10u}.

 \section{Calculations and discussion.}

 Scalar relativistic calculations were performed within the generalized
 relativistic effective core potential (GRECP) model\cite{Titov:99, Petrov:04b,
 Mosyagin:05a, Mosyagin:05b} using the CCSD(T) method (implemented in the 
 {\sc molcas} program package\cite{MOLCAS}) for correlation treatment. 
 The high accuracy and reliability of this approach has been
 demonstrated in similar calculations (see, e.g., \cite{Petrov:09a}).
 The $4f_{5/2}$, $4f_{7/2}$, and $6s_{1/2}$ spinors of the Yb atom have 
 the one-electron energies of -0.54, -0.49, and -0.20~a.u., respectively, 
 and are usually considered as valence ones. The average radii of the 
 $4s_{1/2}$, $4p_{1/2}$, $4p_{3/2}$, $4d_{3/2}$, and $4d_{5/2}$ spinors
 are very close to those of the $4f_{5/2}$ and $4f_{7/2}$ spinors. 
 Therefore, excitations of electrons from the latter spinors will lead 
 to strong relaxation of the former spinors which, therefore, should 
 be considered as the outercore ones for the GRECP generation procedure and 
 ``valence-type'' property calculations\cite{Titov:99}. 
 Thus, we use the GRECP with 42 explicitly treated electrons for each Yb atom.
 In series of preliminary calculations, we have estimated the contributions 
 from correlations with different shells of Yb to the dissociation energy 
 of Yb$_2$ (some of them are presented in \tref{Yb_2}).
 The main contribution is provided by the $6s$ shell whereas
 the contributions from
 the $4f$, $5s$, and $5p$ shells are relavitely small. It is clear that 
 the corresponding contributions from the innermore $4d$, $4p$, $4s$, 
 etc.\ shells will be significantly smaller.
 Thus, we ``freeze'' the $4s,4p,4d$ or $4s,4p,4d,4f,5s,5p$
 shells in 48- and 4-electron calculations, respectively,
 to reduce computational expenses in the present calculations. 
 Generalized correlation basis sets comprising
 $(19,17,7,17,6,1)/[7,8,5,4,3,1]$ functions, basis C (core),
 in the former and $(38,22,24,14,7,1)/[11,10,9,7,5,1]$, basis L (large),
 $(38,22,24,14)/[5,6,4,3]$, basis M (medium, with the $g$
 and $h$ harmonics removed from the previous uncontracted basis set),
 $(38,22,24,14)/[5,5,3,2]$, basis S (small),
 in the latter cases were generated by the procedure developed 
 previously\cite{Mosyagin:00, Mosyagin:01b}.

 Calculations were carried out for internuclear distances from 6 to 14~a.u.
 All our results were rectified using the counterpoise
 corrections (CPC)\cite{Gutowski:86,Liu:89} calculated for the Yb $6s^2$ state
 with one more Yb atom treated as the ghost one. 
 The energies of the rovibrational levels were obtained by solving 
 the rovibrational Schr{\"o}dinger equation with the numerical interatomic 
 potential by the second order finite-difference method\cite{Mitin:98}.
 The stage of calculation of the molecular constant\cite{Mitin:98} 
 begins with fitting the numerical potential curve for the dimer
 by polynomials with the help of the quasi-Hermitian 
 method.  Appropriate derivates of the potential curve at
 the equilibrium point are calculated by recurrence relations.
 Then rovibrational Schr{\"o}dinger equation is solved by the Dunham method 
 to express the Dunham coefficients in terms of these derivates.

\begin{figure}[ht]
   \includegraphics[width=0.6\textwidth]{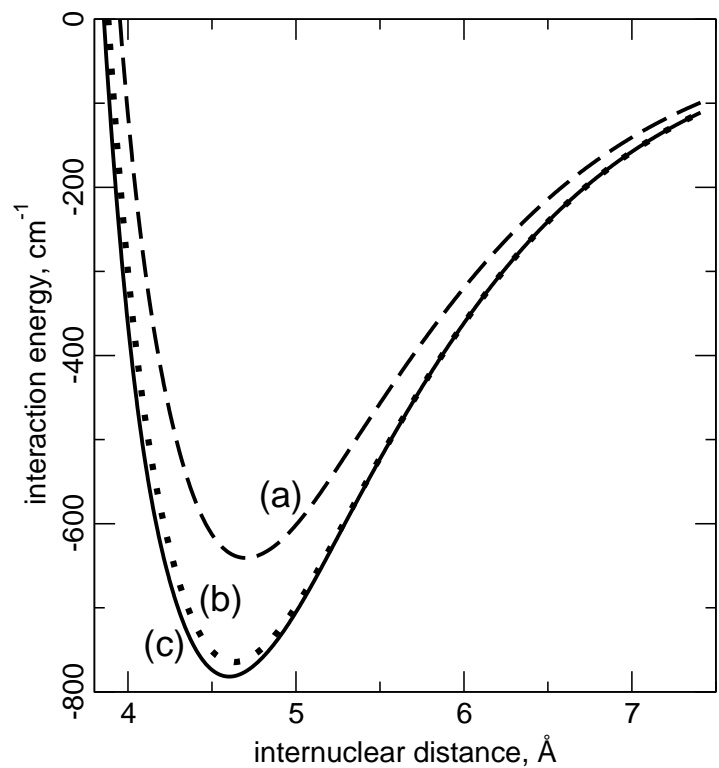}
\caption{ \label{figyb2}
Calculated potential energy functions for the Yb$_2$ 
ground state. Curve (a) corresponding to the computational scheme 
4e-CCSD(T)+OC provides an approximation for all-electron scalar CCSD(T), 
that obtained at the 4e-CCSD(T)+OC+iTQ  level (b) should approach the scalar
relativistic limit whereas curve (c) presents our best full relativistic 
results (4e-CCSD(T)+OC+iTQ+SO).}
\end{figure}

 The $^1\Sigma_g^+$ closed-shell ground state of the Yb$_2$ molecule
 disscociates into two Yb atoms in the $4f^{14} 6s^2 (^1S)$ ground state.
 The computed ground-state potential energy curves for
 the Yb$_2$ molecule are shown in \tref{tabyb2} and \fref{figyb2}; 
 the energies of the lowest rovibrational levels for the 
 ground electronic state are collected in \tref{rovibyb2};
 our estimates for the dissociation energy, equilibrium distance, and 
 main spectroscopic constants are listed in \tref{Yb_2}.
 We started from 4-electron
 scalar relativistic CCSD(T) (denoted as 4e-CCSD(T) below) calculations with 
 rather large basis set L, which gave $D_e=$706~\cm. The negligible
 CPC (0.3~\cm~for dissociation energy) indicates a good quality of the basis
 set used. Subsequent calculations of the effects of the difference between 
 the iterative and non-iterative triple cluster amplitudes 
 (CCSDT-CCSD(T) or contribution from iteration of triples) 
 as well as of quadruple cluster amplitudes
 (these two contributions are denoted further as the iTQ contribution), 
 valence -- outer core correlations (OC), and the spin-orbit interaction (SO)
 described below have shown that the corresponding contributions to the Yb--Yb
 interaction energy are within 15\% 
 (with respect to the final dissociation energy estimate of 786~\cm),
 thus justifying the choice of the 4e-CCSD(T)
 scheme as a good initial approximation. Note that the 4-electron FCI or
 48-electron CCSD(T) calculations with considerably smaller 
 basis sets M or C have given much lower $D_e$ 
 estimates ($D_e=$536 or 353~\cm, correspondingly).
 Thus, the quality of the basis set is of crucial
 importance for accurate calculations of the ytterbium dimer.

 The contribution from the quadruple cluster amplitudes 
 as well as the difference between the iterative and 
 non-iterative triple amplitudes was 
 estimated as the difference between the total energies obtained in 
 the 4e-FCI and 4e-CCSD(T) calculations with basis set M
 for each of the above mentioned
 internuclear distances. This difference was then added to the total energy
 obtained in the 4e-CCSD(T) calculation with basis set L. 
 The derived correction from the iTQ amplitudes 
 to the dissociation energy, 117~\cm, is 15\% 
 with respect to our final value for $D_e$.
 We expect that the contribution from the iTQ amplitudes 
 calculated with basis set M will change very slightly (with respect to 
 the final $D_e$ value, etc.) if this basis set is replaced by basis L.
 It should be noted that the 4e-CCSD energy difference for
 $R_e=100.$~a.u.\ and $R_e=9.$~a.u.\ (the latter is close to the equilibrium
 distance) changes by 229 and 215~\cm~in going from 
 basis set S to M and from basis set M to L,
 respectively. The corresponding contributions from the non-iterative triple
 cluster amplitudes are 94 and 81~\cm, whereas the former contribution from 
 the iTQ amplitudes is only 18~\cm. Thus, extrapolation to the infinite basis
 set limit should only slightly increase the dissociation energy estimate.

 The contribution from the correlations with the $4f$, $5s$, and $5p$ 
 outer-core electrons was estimated as the difference between the total
 energies found in the 48e-CCSD(T) and 4e-CCSD(T) calculations with 
 basis set C for each of the above mentioned internuclear distances. 
 The only difference between these two calculations is the number of 
 correlated electrons, therefore, the lowerings in the total energies
 give the contribution of the OC correlations.
 These lowerings were then added to the 4e-CCSD(T) and 4e-CCSD(T)+iTQ 
 total energies derived above. The 4e-CCSD(T)+OC and 4e-CCSD(T)+iTQ+OC 
 dissociation energy, equilibrium distance and 
 spectroscopic constants were calculated with the obtained potential curves.
 The dissociation energy was decreased by 56~\cm~, whereas the corresponding 
 CPC contribution was about of 100~\cm. It should be noted that 
 the level of the approximations made in 
 calculations\cite{Cao:02,Buchachenko:07} most closely 
 correspond to our 4e-CCSD(T)+OC approximation (see \tref{Yb_2}).  
 The difference (of 82--98~cm$^{-1}$ for $D_e$) between the results of the
 above calculations can be assigned to the insufficient flexibility of
 the basis set functions from the ``outer-core'' region in
 \cite{Cao:02,Buchachenko:07}. Unfortunately, calculations for 
 4 correlated electrons, which are necessary to check this conjecture, 
 were not reported in the cited works.

 The effect of spin-dependent (effective spin-orbit) interactions was
 taken from Ref.~\cite{Zaitsevskii:09u} as the difference between
 the ground-state potential curves obtained in
 two-component relativistic DFT calculations\cite{NWChem47} with full RECPs 
 and in scalar relativistic DFT calculations with spin-averaged RECPs.
 The details of the employed procedure can be found
 elsewhere\cite{Zaitsevskii:06a}.
 We only note that effects of electronic correlations are taken into account
 within DFT, so the additivity of the correlations and spin-orbit effects is
 irrelevant to (is not exploited in) our present study.
 An uncontracted Gaussian basis
 set\cite{Zaitsevskii:09u} $(10s11p8d9f4g)$ was
 used to expand auxiliary one-electron spinors in the
 Kohn-Sham scheme. Two generalized gradient approximations for
 exchange-correlation functionals were employed, 
 a rather universal Perdew-Burke-Erzernhof
 (PBE) model\cite{Perdew:96} and the Perdew-Wang approximation
 (PW91\cite{Perdew:92b}), which is often
 believed to be particularly well suited for the description of van-der-Waals 
 bonds\cite{Tsuzuki:01}.

 Scalar relativitic DFT dissociation energy is about 1.6 times higher 
 than that of the corresponding {\it ab initio} 4e-CCSD(T)+iTQ+OC approximation.
 Nevertheless, we believe that the spin-orbit contribution is rather
 correctly extracted from the DFT calculations because 
 the spin-orbit interaction is described by the the one-electron operator
 and the one-electron parts are the same for Kohn-Sham and Schr{\"o}dinger 
 (including Hartree-Fock) Hamiltonians.
 Despite the potential energy functions obtained in the relativistic DFT 
 calculations with PBE and PW91 functionals
 are slightly different in shape (dissociation
 energy estimates in two-component calculations are 1238 and 1298~\cm,
 respectively), the corresponding spin-orbit corrections to bond energies as
 functions of the internuclear separation almost coincide. The addition of
 these corrections to the results of accurate scalar relativistic calculations
 has increased $D_e$ by 19~\cm.

\section{Conclusions.}

 We predict the exact $D_e$ and $w_e$ to be slightly higher than
 786~\cm~and 24.1~\cm~and the exact $R_e$ to be slightly lower than
 4.582~\AA~, because all contributions (taken into account in our calculations
 with a good accuracy) except for the OC correlations change 
 these constants corresponingly.
 We expect that the reported estimates of the 
 dissociation energy, equilibrium distance and spectroscopic
 constants of Yb$_2$ obtained by the
 CCSD(T) technique with very extensive basis sets and the incorporation of
 corrections for higher-order cluster amplitudes and spin-orbit interactions 
 are the most reliable up to date. Our analysis has revealed a 
 non-negligible role of quadruple amplitudes as well as the
 significant contribution from iteration of triple amplitudes
 in the cluster expansion (which were not taken into account in 
 \cite{Dolg:92a,Wang:98b,Cao:02,Buchachenko:07}) and small 
 but non-negligible contributions from spin-dependent relativistic effects.

\ack
 We are grateful to S.Kotochigova for initiating this work 
 and to A.V.Zaitsevskii and E.A.Rykova for the DFT calculations of Yb$_2$.
 The present work was supported by RFBR grants 09--03--01034, 09--03--00655,
 and, partially, by RFBR grant 07--03--01139.

\section*{References}
\bibliographystyle{unsrt}

\bibliography{bib/JournAbbr,bib/Titov,bib/TitovLib,bib/Kaldor,bib/Isaev,bib/TitovAbs,bib/MosyaginLib}

\Tables
\begin{table}
\caption{\label{tabyb2}
Potential energy functions for the Yb$_2$ ground state
calculated with the help of the GRECP and different correlation methods. 
Internuclear distances $R$ and total energy lowerings $E(R)-E(\infty)$ 
are in a.u.
}
\begin{indented}
\item[]
\begin{tabular}{@{}rrrr}
\br
 $R$ & \multicolumn{3}{c}{$E(R)-E(\infty)$}                 \\
\cline{2-4}                                                   
     & 4e-CCSD(T)+OC & 4e-CCSD(T)+OC+iTQ & 4e-CCSD(T)+OC+iTQ+SO \\
\mr                                                           
   6 &    0.02783884 &      0.02661730 &         0.02244608 \\
   7 &    0.00432099 &      0.00334087 &         0.00280762 \\
   8 &   -0.00206658 &     -0.00281218 &        -0.00299103 \\
   9 &   -0.00290975 &     -0.00343593 &        -0.00349171 \\
  10 &   -0.00236798 &     -0.00271825 &        -0.00273208 \\
  11 &   -0.00166549 &     -0.00188952 &        -0.00189083 \\
  12 &   -0.00110258 &     -0.00124272 &        -0.00124178 \\
  13 &   -0.00070971 &     -0.00079694 &        -0.00079630 \\
  14 &   -0.00045218 &     -0.00050694 &        -0.00050653 \\
 100 &    0.00000000 &      0.00000000 &         0.00000000 \\
\br 
\end{tabular}
\end{indented}
\end{table}

\begin{table}
\caption{\label{rovibyb2}
The energies of the lowest rovibrational levels for the $^1\Sigma_g^+$ 
ground state of the $^{171}$Yb$_2$
molecule from the GRECP/4e-CCSD(T)+iTQ+OC+SO
potential curve are in \cm. J and v are the rotational 
and vibrational quantum numbers.
}
\begin{indented}
\item[]
\begin{tabular}{@{}rrrrrrrrrrr}
\br
     &  J=0   &  J=1   &  J=2   &  J=3   & J=4    &  J=5   &  J=6   &  J=7   &  J=8   \\
\mr                                                                                    
 v=0 &  12.26 &  12.28 &  12.31 &  12.37 &  12.45 &  12.54 &  12.65 &  12.78 &  12.93 \\
 v=1 &  36.23 &  36.25 &  36.29 &  36.34 &  36.42 &  36.51 &  36.62 &  36.75 &  36.90 \\
 v=2 &  59.45 &  59.47 &  59.51 &  59.56 &  59.64 &  59.73 &  59.84 &  59.97 &  60.11 \\
 v=3 &  82.15 &  82.17 &  82.21 &  82.26 &  82.34 &  82.43 &  82.54 &  82.66 &  82.81 \\
 v=4 & 104.44 & 104.46 & 104.49 & 104.55 & 104.62 & 104.71 & 104.82 & 104.94 & 105.09 \\
 v=5 & 126.33 & 126.35 & 126.39 & 126.44 & 126.51 & 126.60 & 126.71 & 126.83 & 126.97 \\
 v=6 & 147.85 & 147.87 & 147.90 & 147.96 & 148.03 & 148.12 & 148.22 & 148.35 & 148.49 \\
 v=7 & 169.00 & 169.02 & 169.05 & 169.10 & 169.17 & 169.26 & 169.37 & 169.49 & 169.63 \\
 v=8 & 189.77 & 189.79 & 189.82 & 189.87 & 189.94 & 190.03 & 190.14 & 190.26 & 190.40 \\
\br 
\end{tabular}
\end{indented}
\end{table}

\begin{table}
\caption{\label{Yb_2}
The dissociation energy, equilibrium distance, and 
spectroscopic constants of the $^1\Sigma_g^+$ ground state
of the $^{171}$Yb$_2$ molecule. $R_e$ is in \AA, 
$B_e$ in $10^{-3}$~\cm, $\alpha_e$ in $10^{-5}$~\cm, $Y_{02}$ in $10^{-9}$~\cm, 
and other values in \cm.
}
\begin{indented}
\item[]
\begin{tabular}{@{}lcccccccc}
\br
 Method                              & $D_e$ & $R_e$ & $w_e$ & $D_0^0$ & $B_e$ & $w_e x_e$ & $\alpha_e$ & $-Y_{02}$ \\
\mr                                                                                                                          
\multicolumn{9}{c}{Present GRECP calculations:}                                                                                  \\
 4e-CCSD(T)                          & 706   & 4.767 & 22.9  &  694    &  8.67 &  0.20     &  7.5       & 5.0       \\
 4e-CCSD(T)+OC                       & 642   & 4.683 & 21.5  &  631    &  8.99 &  0.19     &  8.3       & 6.3       \\
 4e-CCSD(T)+iTQ                        & 823   & 4.708 & 24.7  &  811    &  8.89 &  0.20     &  7.1       & 4.6       \\
 4e-CCSD(T)+iTQ+OC                     & 767   & 4.615 & 23.5  &  756    &  9.25 &  0.19     &  7.8       & 5.8       \\
 4e-CCSD(T)+iTQ+OC                     &       &       &       &         &       &           &            &           \\
 ~~~~~~+SO                             & 786   & 4.582 & 24.1  &  774    &  9.39 &  0.23     &  8.3       & 5.7       \\
\multicolumn{9}{c}{Previous calculations:}                                                                                 \\
 42e-GRECP/                          &       &       &       &         &       &           &            &           \\
 ~~~~~~DFT(PW91)\cite{Zaitsevskii:09u}  & 1261  & 4.274 & 33.2  & 1244    & 10.79 &  0.19     &  6.3       & 4.5       \\
 10e-PP/CISD\cite{Dolg:92a}          & 400   & 5.308 & 13    &         &       &           &            &           \\
 10e-PP/20e-CCSD(T)\cite{Wang:98b}   & 470   & 4.861 & 18    &         &       &           &            &           \\
 42e-PP/CCSD(T)\cite{Cao:02}         & 740   & 4.549 & 25    &         &       &           &            &           \\
 42e-PP/CCSD(T)\cite{Buchachenko:07} & 724   & 4.472 &       &         &       &           &            &           \\
\br 
\end{tabular}
\end{indented}
\end{table}

\Figures
                                                                                
%Figure 1, N.S.Mosyagin, {\it et al.}, Journal of Physics B

\end{document}